\documentclass[twocolumn, amssymb, nobibnotes, aps, prb, superscriptaddress]{revtex4-2}

\newcommand{\be}{\begin{equation}}
\newcommand{\ee}{\end{equation}}

\usepackage{amsmath, amsfonts, amstext, amssymb}
\usepackage[english]{babel}
\usepackage[latin1]{inputenc}
\usepackage{subfigure}
\usepackage[thinlines]{easytable}
\usepackage{siunitx}
\usepackage{exscale}
\usepackage{tabulary}
\usepackage{ae}
\usepackage{graphicx, wrapfig}
\usepackage{enumerate}
\usepackage{wrapfig}
\usepackage{placeins}
\usepackage[dvipsnames]{xcolor}

\usepackage[normalem]{ulem}

\newcommand{\revisionKB}[1]{\textcolor{black}{#1}}
\newcommand{\revisionKBR}[1]{\textcolor{black}{#1}}
\newcommand{\sample}{$\alpha$-MnTe}

\begin{document}

\title{Femtosecond phononic coupling to both spins and charges in a room temperature antiferromagnetic semiconductor}%

\author{D. Bossini} \email[]{davide.bossini@uni-konstanz.de}
\affiliation{Department of Physics and Center for Applied Photonics, University of Konstanz, D-78457 Konstanz, Germany.}
\affiliation{Department of Physics, TU Dortmund University, Otto-Hahn Stra\ss{}e 4, 44227 Dortmund, Germany}
\author{S. Dal Conte}
\affiliation{Dipartimento di Fisica, Politecnico di Milano, Piazza Leonardo da Vinci 32, Milano, Italy}
\affiliation{Istituto di Fotonica e Nanotecnologie, Consiglio Nazionale delle Ricerche, Piazza Leonardo da Vinci 32, Milano, Italy}		
\author{M. Terschanski}
\affiliation{Department of Physics, TU Dortmund University, Otto-Hahn Stra\ss{}e 4, 44227 Dortmund, Germany}
\author{G. Springholz}
\affiliation{Institute of Semiconductor and Solid State Physics, Johannes Kepler University Linz, Altenbergerstr. 69, 4040 Linz, Austria}
\author{A. Bonanni}
\affiliation{Institute of Semiconductor and Solid State Physics, Johannes Kepler University Linz, Altenbergerstr. 69, 4040 Linz, Austria}
\author{K. Deltenre}
\affiliation{Department of Physics, TU Dortmund University, Otto-Hahn Stra\ss{}e 4, 44227 Dortmund, Germany}
\author{F. Anders}
\affiliation{Department of Physics, TU Dortmund University, Otto-Hahn Stra\ss{}e 4, 44227 Dortmund, Germany}
\author{G.S. Uhrig}
\affiliation{Department of Physics, TU Dortmund University, Otto-Hahn Stra\ss{}e 4, 44227 Dortmund, Germany}
\author{G. Cerullo}
\affiliation{Dipartimento di Fisica, Politecnico di Milano, Piazza Leonardo da Vinci 32, Milano, Italy}
\affiliation{Istituto di Fotonica e Nanotecnologie, Consiglio Nazionale delle Ricerche, Piazza Leonardo da Vinci 32, Milano, Italy}
\author{M. Cinchetti}
\affiliation{Department of Physics, TU Dortmund University, Otto-Hahn Stra\ss{}e 4, 44227 Dortmund, Germany}
\date{\today}%

 \begin{abstract}

Spintronics is postulated on the possibility to employ the magnetic degree of freedom of electrons for computation and couple it to charges. In this view, the combination of the high-frequency of spin manipulations offered by antiferromagnets, with the wide tunability of the electronic properties peculiar of semiconductors provides a promising and intriguing platform. Here we explore this scenario in \sample , which is a semiconductor antiferromagnetically ordered at room temperature. Relying on a Raman mechanism and femtosecond laser pulses, we drive degenerate modes of coherent optical phonons, which modulate the chemical bonds involved in the super-exchange interaction. The spectrally-resolved measurements of the transient reflectivity reveal a coherent modulation of the band-gap at the frequency of 5.3 THz. The detection of the rotation of the polarisation, typically associated with magneto-optical effects, shows coherent and incoherent contributions. Modelling how the ionic motion induced by the phonons affects the exchange interaction in the material, we calculate the photoinduced THz spin dynamics: the results predict both a coherent and incoherent response, the latter of which is consistent with the experimental observation. Our work demonstrates that the same phonon modes modulate both the charge and magnetic degree of freedom, suggesting the resonant pumping of phonons as a viable way to link spin and charge dynamics even in nonlinear regimes.
%
 \end{abstract}


\maketitle

\pagenumbering{arabic}

\section{Introduction}

One of the overarching goals of solid state physics in the last thirty years has been to establish magnetic semiconductors as a platform for dissipationless spintronics~\cite{Furdyna:1988bt,10.1038/nmat2898,10.1038/nphys551}. The main hindrance to this breakthrough is the difficulty to synthesize a room-temperature ferromagnetic semiconductor. Recently, the spintronic capabilities of metallic and dielectric antiferromagnets (AFs) have been deeply investigated~\cite{Jungwirth:2018if,Jungwirth2016,Baltz:2018iv}, confirming the feasibility of spin manipulation even at extreme time-scales~\cite{Bossini:2016iv,Nemec:2018gw}. Here, we measure the time-dependent optical properties of the room temperature antiferromagnetic semiconductor MnTe. Ultrafast spintronics in semiconductors requires the ability to couple the magnetic and electronic degrees of freedoms, the latter being strongly defined by the band-gap, in the femtosecond regime. We demonstrate that the displacive excitation~\cite{Zeiger1992,Merlin1997a} of an optical phonon modulates coherently the band-gap at a 5.3 THz frequency. Our data show that the same lattice mode couples also to spins. We derive a model revealing the nature of this phenomenon and predicting the spin dynamics induced by this coupling, whose signatures are observed in the time-resolved magneto-optical response. Given the simultaneous dynamical coupling to both spins and charges, the selective resonant excitation of optical phonons emerges as a tool to advance towards the long sought-after realization of efficient and controllable semiconductor spintronics.

This paper is organized as follows. We introduce in Sec. \ref{sec:Methods} the material here investigated and the experimental methods, discussing both the set-up and the magneto-optical effects allowed in our detection geometry. In Sec. \ref{sec:BandGap}, we present the time- and spectrally-resolved measurements of the reflectivity of \sample. The optical dynamics detected via single-colour experiments, revealing the generation of coherent phonons, is shown and discussed in Sec.\ref{sec:1ColOp}. In Sec. \ref{sec:theory} our model of the spin dynamics induced by the optically-generated phonons is reported. The measurement of the magneto-optical dynamics is the subject of Sec. \ref{sec:1ColMO}. We summarize the conclusions and formulate the perspectives of our work in Sec. \ref{sec:Concl}. 

\section{Methods and Materials}
\label{sec:Methods}

\subsection{Material}

\begin{figure}[h!]
	\centering
	\includegraphics[width=\columnwidth]{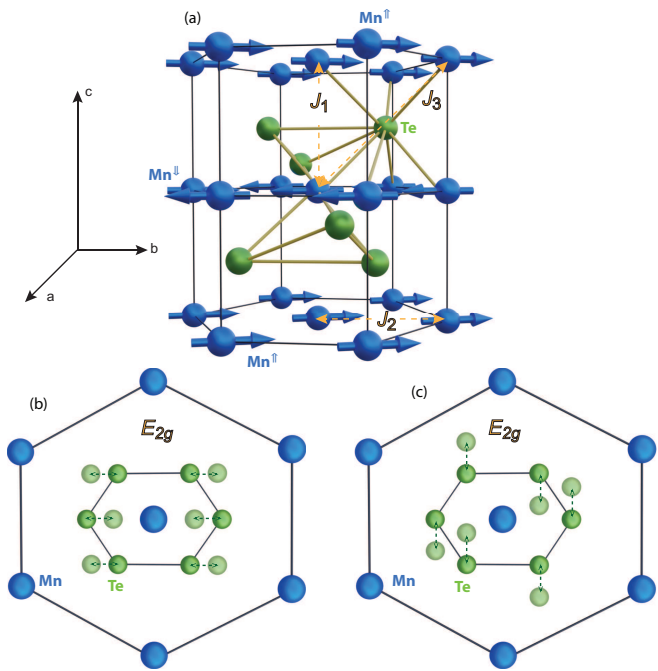}
	\caption{\footnotesize{Structure and lattice modes of \sample\ . (a) Crystallographic and magnetic structure of \sample\ . The three super-exchange couplings are shown by the yellow dashed arrows~\cite{Szuszkiewicz:2006fz}: $J_1$ is the dominating antiferromagnetic super-exchange interaction, $J_2 $ is a weaker ferromagnetic coupling between Mn-atoms within the ferromagnetic hexagonal planes, the Te-mediated $J_3$ is the weakest interaction and it is antiferromagnetic, involving 12 next-nearest neighbour sites. (b) and (c) Displacement of Te atoms induced by the degenerate $E_{2g}$ modes. The identification of the atomic displacement is done by comparing \sample\ to the isomorphic MgB$_2$~\cite{Rzhevskii2009}.}}
\label{fig:Fig1}
\end{figure}

We investigate hexagonal manganese telluride (\sample). This material is a correlated magnetic semiconductor comprising two antiferromagnetically coupled magnetic sublattices (Mn$^{\Uparrow,\Downarrow}$ in Fig. \ref{fig:Fig1}(a)), which originate a long-range order above room temperature (N\'eel temperature $T_{\mbox{\tiny{N}}}\approx$ 307 K\cite{Kriegner:2016dj,Mu:2019jg,Szuszkiewicz:2006fz}). The reported band-gap $E_{\text{gap}}$ $\approx (1.27-1.46)$ eV is indirect~\cite{Kriegner:2016dj}. \sample\ crystallizes in the NiAs structure\cite{Mobasser:1985bt} shown in Fig. \ref{fig:Fig1}(a) and has a centrosymmetric point group ($6/mmm$), in which the collective ionic vibrations of the Te atoms displayed in Fig. \ref{fig:Fig1}(b)-(c) represent Raman-active optical phonons. As such, the nearest neighbours Te atoms oscillate out-of-phase as shown in Fig. \ref{fig:Fig1}(b)-(c). Considering that Mn- and Te-ions are charged, the ionic oscillations represented in Fig. \ref{fig:Fig1}(b)-(c) induce transient dipole moments and, thus, local electric fields during the lattice motion. Our specimen of $\alpha$-MnTe is grown by molecular beam epitaxy on a SrF$_2$(111) substrate using elemental Mn and Te sources and substrates temperatures in the range (370-450) $^{\circ}$C. Two-dimensional growth is achieved in both cases as inferred from the streaked reflection high energy electron diffraction (RHEED) patterns observed during the growth. The epitaxial relation between the MnTe overlayer and the substrate is (0001)[10$\bar{1}$0]$_{\mbox{\tiny{MnTe}}}$|| (111)[11$\bar{2}$]$_{\mbox{\tiny{sub}}}$. 

\subsection{Pump-probe set-up}
For the pump-probe experiments a regeneratively amplified mode-locked Ti:Sapphire laser, providing 100-fs, 500-$\mu$J pulses at 800 nm and 2 kHz repetition rate is used. For broadband transient reflectivity measurements (Fig. \ref{fig:Fig2}) the \sample\ sample is photoexcited by a broadband pulse centred at 2.4 eV generated by a home-built non-collinear optical parametric amplifier (NOPA). The pulse is temporally compressed to nearly 15 femtoseconds by a pair of custom-made chirped mirrors. The transient optical response of the sample is measured by a white-light probe generated in a 2 mm-thick sapphire plate. After the generation, the fundamental wavelength beam is removed by a short-pass filter. The rms fluctuation of white light is around 0.2$\%$. The pump and probe spot sizes at the sample position are 100 $\mu$m and 70 $\mu$m, respectively. The reflected probe is dispersed by a grating and detected by a Si CCD camera with 532 pixels. The pump pulse is modulated at 1 kHz by a mechanical chopper, resulting in a detection sensitivity of the order of 10$^{-4}$ with an integration time of 2 s.

For the transient transmissivity and polarization rotation measurements (Fig. \ref{fig:Fig5}) the probe is generated by a second NOPA, filtered around 1.72 eV and collected by a balanced photodetector, consisting of two photodiodes followed by a differential amplifier. The Wollaston prism separates the two orthogonally polarized components of the probe beam (Fig. \ref{fig:Fig4}(a)). We rotate the Wollaston  prism in order to equalize the probe intensities on the two photodiodes. The pump-induced imbalance of the signal is measured by a lock-in amplifier which is locked to the modulation frequency of the pump beam (i.e. 1 kHz). The sample has been placed in a cryostat, allowing to cool it down to liquid nitrogen temperature.

\subsection{Magneto-optical detection}

\begin{figure}[h!]
	\centering
	\includegraphics[width=\columnwidth]{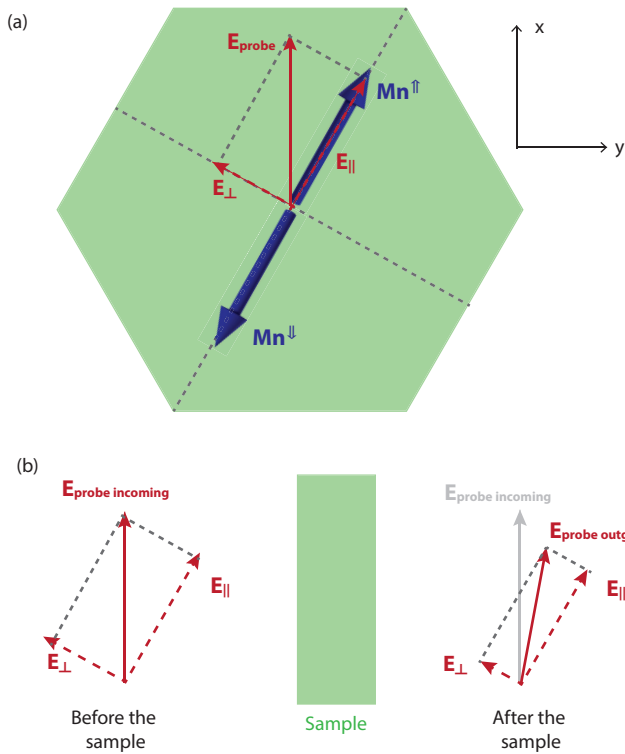}
	\caption{\footnotesize{{\bfseries{MLD in $\alpha$-MnTe.}}(a) Orientation of the N\'eel vector for one of the 3 possible antiferromagnetic domains~\cite{Kriegner:2016dj}. The electric field of the probe beam has components along both the direction of spins and the direction perpendicular to them. (b) Rotation of the polarization induced by magnetic linear dichroism for a probe beam propagating through the sample.}}
	\label{fig:MLDsk}
\end{figure}
	
Time-resolved spin dynamics in antiferromagnets can be typically detected by means of quadratic magneto-optical effects affecting the probe beam. The dielectric tensor describing the interaction between \sample\ and the probe beam, propagating along the z-axis (reference frame in Fig. \ref{fig:MLDsk}) is given by

\begin{equation}
	\epsilon =
	\begin{pmatrix}
 	 \epsilon_{xx} & \epsilon_{xy} & 0 \\
 	 \epsilon_{yx} & \epsilon_{yy} & 0 \\
 	 0 & 0 & 0
	\end{pmatrix}.
\end{equation}

\noindent In an antiferromagnetic material the symmetric components can be expanded~\cite{Landau1981} as

\begin{equation}
\epsilon_{ii} = \epsilon_{ii}^0 + \chi_{kl}L_kL_l + \xi_{klmn}L_kL_lL_mL_n +... 
\label{eq:ep}	
\end{equation}

\noindent where $L_i$ are the components of the N\'eel vector. The first term is spin-independent and describes the refraction (real part) and the optical absorption (imaginary part) of the material. The following terms contain even powers of spin multiplied by material-dependent tensors. The second term describes quadratic-magneto-optical effects, which are common in antiferromagnets~\cite{Bossini:2017bo,Nemec:2018gw}. We report here a symmetry analysis of \sample\ in the experimental geometry shown in Fig. {\ref{fig:MLDsk}}. The hexagonal \sample\ belongs to the $6/mmm$ point group, therefore the light-matter interaction originating quadratic magneto-optical effects in general is described by the following contribution to the thermodynamic potential

\begin{equation}
	\Phi = \chi_{ijkl}E_iE_jL_kL_l,
	\label{eq:MO}	
\end{equation}

\noindent where $i,j,k,l$ are cartesian coordinates. In Eq. (\ref{eq:MO}) $\chi_{ijkl}$ is a fourth-rank polar tensor, $E_{i,j}$ are the components of the electric field of the pump beam and $L_{k,l}$ are components of the N\'eel vector. In the reference frame shown in Fig. (\ref{fig:MLDsk}), the electric field of the probe beam was linearly polarized along the the x-axis and propagating along the z-axis. The N\'eel vector has components along both the x- and y-axis (Fig. (\ref{fig:MLDsk})), given the domain structure of an hexagonal antiferromagnet~\cite{Kriegner:2016dj}. Considering the specific case of magnetic linear dichroism, the antiferromagnetic vector has to be considered quadratically~\cite{Ferre1984}, so we obtain

\begin{equation}
	\begin{aligned}
		\Phi &= \chi_{iijj}E_iE_iL_jL_j =\\ 
		& = \chi_{xxxx}E_xE_xL_xL_x + \chi_{xxyy}E_xE_xL_yL_y.
	\end{aligned}
\end{equation}

\noindent The components of $\chi$ satisfy the symmetries of the $6/mmm$ point group~\cite{Birss1966}, namely $\chi_{xxyy} = 1/3 \cdot \chi_{xxxx} \neq 0$ showing that magnetic linear dichroism is allowed by symmetry in $\alpha-$Mnte.

Figure \ref{fig:MLDsk}(a) schematically represents the magnetic linear dichroism in $\alpha$-MnTe in our experimental geometry. A Mn atom is located in correspondence of each corner of the hexagon. Considering one of the three in-plane easy axes of this material as the direction along which spins are aligned, we can identify a parallel and a perpendicular direction to the antiferromagnetic vector. The probe beam has electric field parallel to the x-axis, which means that it has both a non-zero parallel and a non-zero perpendicular component with respect to the direction of the N\'eel vector. As a consequence, it experiences the absorption coefficients along these two directions, which differs because of the magnetic contributions to the dielectric tensor. In the case of a dissipative light-matter interaction regime, these terms and $\Im{\epsilon_{ii}^0}$ can affect the rotation of the polarization via linear dichroism, which is magnetic if spin-dependent terms are involved: magnetic linear dichroism (MLD) induces thus a rotation of the polarization quadratically proportional to spins. Therefore the ratio $E_{\parallel}/E_{\perp}$ for the probe beam transmitted through the sample is different from the value of this ratio for the probe beam impinging on the sample. This change in the $E_{\parallel}/E_{\perp}$ ratio induces a rotation of the electric field of the probe beam and, thus, of the polarization (Fig. \ref{fig:MLDsk}(b)). We note that in the reference frame displayed in Fig. \ref{fig:MLDsk}(a), a linear combination of $L_x$ and $L_y$ expresses the purely longitudinal spin dynamics. As reported in the literature~\cite{Bossini2014,Bossini:2017bo,10.1038/nphoton.2016.255,Nemec:2018gw}, this magneto-optical effect can be employed in pump-probe experiments to detect the transient longitudinal demagnetization of the sublattices in an AF.
 
We now move on discussing the measurements of the transient reflectivity performed with both temporal and spectral resolution.
 
\section{Time- and spectrally-resolved optical measurements}
\label{sec:BandGap}

\begin{figure}[h]
	\centering
	\includegraphics[width=\columnwidth]{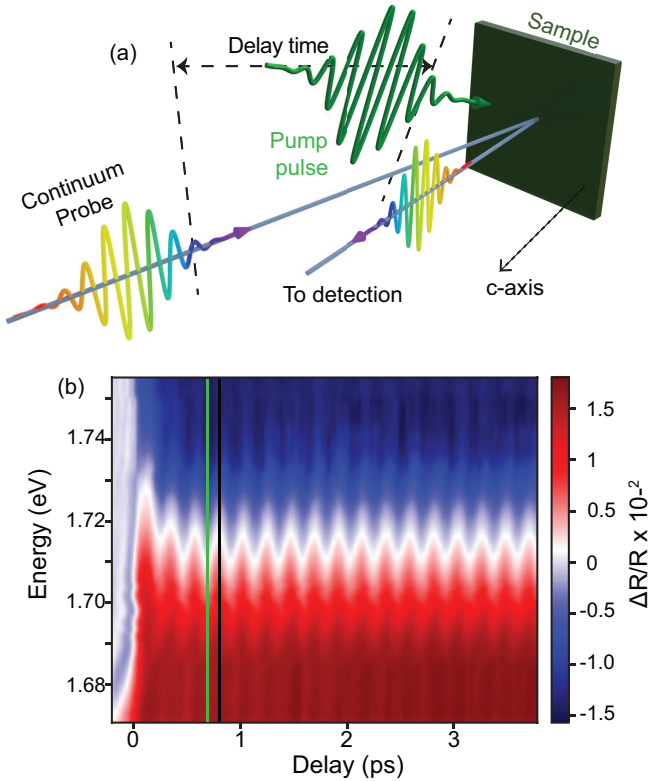}
	\caption{\footnotesize{Time- and spectrally-resolved femtosecond dynamics of the band-gap of MnTe. (a) The pump beam with a central photon energy of 2.4 eV impinges on the sample, triggering dynamics of the reflectivity. The transient reflectivity in the 1.65 eV - 2.76 eV spectral range is monitored by white light continuum pulses, which are delayed in a controlled and continuously variable way from the excitation pulses. (b) Spectrally- and time-resolved dynamics of the reflectivity of \sample\ triggered by a linearly polarized pump beam. The fluence is set to 60 $\mu$J/cm$^2$ and the temperature of the sample is 77 K. Oscillations at the frequency of 5.3 THz are visible in the spectral band centred at 1.71 eV, 50 meV wide. The two vertical lines show two selected pump-probe delays corresponding to a maximum and minimum of the oscillations of $E_{\text{gap}}$.}}
	\label{fig:Fig2}
\end{figure}

We perform a femtosecond pump-probe experiment, in which the transient reflectivity ($\Delta R/R$) of \sample\ is detected in the (1.65 - 2.76) eV spectral range (Fig. \ref{fig:Fig2}(a)). The combination of high temporal and spectral resolution of this approach allows to fully track the photo-driven transient evolution of $E_{\text{gap}}$. The colormap in Fig. \ref{fig:Fig2}(b) shows two spectral regions with a different sign of the signal: negative above 1.71 eV and positive below. Interestingly, in a spectral range centred at 1.71 eV, which corresponds to the rising edge of the band-gap~\cite{FerrerRoca:2000ku,Bossini:2020fs}, pronounced oscillations at the frequency of 5.3 THz are visible. This value is consistent with the frequency of two degenerate $E_{2g}$ phonon modes observed in the Raman spectrum of \sample\ ~\cite{Zhang:2020jl}. These lattice modes imply oscillations of Te atoms, as displayed in Fig. \ref{fig:Fig1}(b)-(c). 

\begin{figure}[h]
	\centering
	\includegraphics[width=\columnwidth]{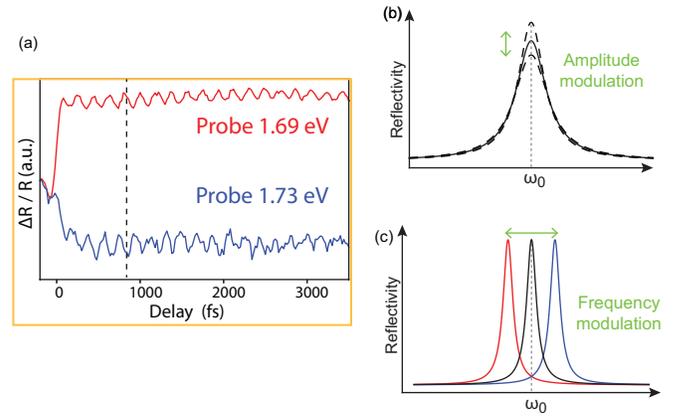}
	\caption{\footnotesize{Coherent modulation of the band-gap. (a) The oscillations of $\Delta R/R$ for the selected spectral components of the supercontinuum probe are out of phase by $\pi$. (b) Amplitude modulation of a Lorentz oscillator: the spectral components at both sides of the resonance frequency ($\omega_0$) oscillate in phase. (c) Frequency modulation of a Lorentz oscillator: the spectral components at the blue and red sides of the resonance frequency ($\omega_0$) oscillate out of phase by $\pi$. This modulation is consistent with the data shown in Fig. \ref{fig:Fig2}(b).}}
	\label{fig:Fig3}
\end{figure}

Selecting one probe photon-energy just above and just below the central value of 1.71 eV (Fig. \ref{fig:Fig3}(a)), we observe that the oscillations of the reflectivity are out-of-phase by $\pi$. To understand the phase of the oscillations we model the band-gap of \sample\ by a simple Lorentz oscillator, modulated by the optically-activated phonon. A modulation of the amplitude (Fig. \ref{fig:Fig3}(b)) triggers oscillations of all the spectral components in phase with each other. In contrast, if the frequency ($\omega_0$ in Fig. \ref{fig:Fig3}(b)-(c)) is modulated (Fig. \ref{fig:Fig3}(c)), the spectral components on the higher-energy side of $\omega_0$ oscillate with a phase difference of $\pi$ with respect to the components on the lower-energy side of the central frequency. This latter scenario is consistent with our data and points to the conclusion that the band-gap is coherently modulated at the frequency of 5.3 THz. The degenerate $E_{2g}$ lattice modes are optical, so as already mentioned in Sec. \ref{sec:Methods}, they induce a transient local electric field, which can shift the spectral lines via a Stark-like effect. This phenomenon has been experimentally already observed in both direct (GaAs) and indirect (Si) band-gap semiconductors\cite{Hase:2013ec}. 


\begin{table}
	\begin{TAB}(r,1cm,1.5cm)[5pt]{|c|c|c|}{|c|c|c|c|}
	\textbf{Material} & \textbf{Fuence} & $\mathbf{\Delta R/R} \ $ \\
	Si & \quad 300 $\mu$J /cm$^2$ \quad & $10^{-5} \ $ \\
	GaAs & \quad 170 $\mu$J /cm$^2$ \quad & 10$^{-4} \ $ \\
	\sample & \quad 60 $\mu$J /cm$^2$ \quad & 10$^{-2} \ $\\
	\end{TAB}
\caption{Comparison efficiency. The data for Si and GaAs are taken from the literature\cite{Hase:2013ec}.}
\label{tb:Comparison}
\end{table}

A quantitative comparison of the efficiency of the process is reported in Table \ref{tb:Comparison}. The data in Fig. \ref{fig:Fig2}(b) are obtained by pumping \sample\ with mere 60 $\mu$J/cm$^2$ and the $\Delta R/R$ modulation observed ($1.5 \cdot 10^{-2}$ ) is two orders of magnitude higher than in GaAs, despite the fluence being three times smaller. We thus conclude that the lattice-$E_{\text{gap}}$ coupling in \sample\ induces a much stronger transient modulation of the reflectivity in comparison with the samples investigated in the literature\cite{Hase:2013ec}. Ascertaining whether this difference is due to the cross section of the displacive excitation mechanism or to the coupling between the lattice mode and the profile of the band structure is beyond the scope of the current manuscript. Moreover, we are interested in estimating the  value of the band-gap modulation observed in Fig. \ref{fig:Fig2}(b). The intrinsic temporal chirp of the supercontinuum due to the group velocity dispersion must be taken into account~\cite{Manzoni}. This effect is visible at the femtosecond time-scale. The chirp is numerically corrected by separately analysing the transient reflectivity for each wavelength of the supercontinuum probe beam. In particular, the first derivative of each pump-probe scan is calculated. Then the maximum of the first derivative is defined as zero delay time. 

\begin{figure}[h!]
	\centering
	\includegraphics[width=\columnwidth]{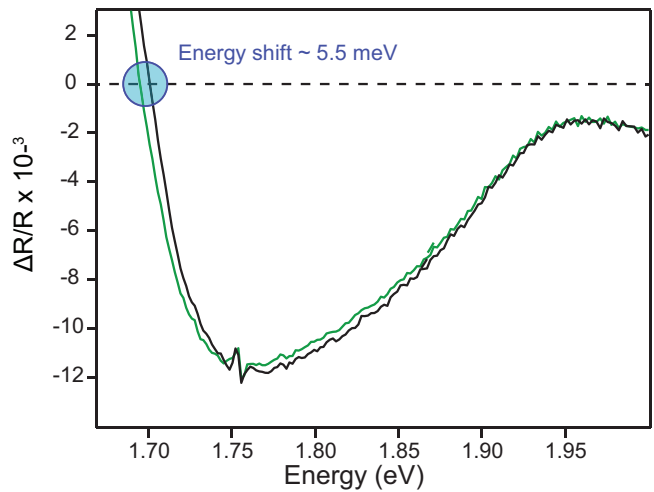}
	\caption{\footnotesize{{\bfseries{Extrapolation $\Delta E_{\text{gap}}$.}} Transient reflectivity for the selected pump-probe delays. The value of the energy at which the zero-value (i.e. max or minimum of the oscillation in Fig. \ref{fig:Fig2}(b)) is crossed is considered for both traces. The difference of these two energy values is an estimation of the amplitude of the band-gap modulation.}}
	\label{fig:Extrap}
\end{figure}
	
In Fig. \ref{fig:Fig2}(b) the white stripe representing the phonon oscillation corresponds to a vanishing value of the transient reflectivity. Hence we consider two pump-probe delays, at which a maximum and a minimum of the phonon are observed (vertical lines in Fig. \ref{fig:Fig2}(b)). The procedure employed to estimate the amplitude of the band-gap modulation is shown in Fig. \ref{fig:Extrap}. The reflectivity spectra corresponding to the black and green lines displayed in Fig. \ref{fig:Fig2}(b) are plotted, highlighting the energy region corresponding to a zero-value of the reflectivity (blue circle). The difference in the values of the energy at which the reflectivity vanishes is taken as a direct estimation of the amplitude of the band-gap modulation. The results is $\Delta E_{\text{gap}} \approx 5.5$ meV $\approx 4\cdot 10^{-3}$ of the original value.

The value of the energy gap and the sublattice magnetizations in \sample\ are coupled in equilibrium, as $E_{\text{gap}}$ is blue-shifted in the AF phase by an amount proportional to the modulus of the sublattice magnetization (Mn $^{\Uparrow,\Downarrow}$ in Fig. \ref{fig:Fig1}(a))~\cite{FerrerRoca:2000ku,Bossini:2020fs}. The existence of an out-of-equilibrium coupling between these quantities on the femtosecond time-scale can be experimentally assessed by comparing the optical and magneto-optical dynamics in the paramagnetic (at T = 320 K) and antiferromagnetic (at T = 77 K) phases. Such comparison requires single-colour measurements 

\section{Single-color optical pump-probe measurements}
\label{sec:1ColOp}

\begin{figure}[h]
	\centering
	\includegraphics[width=\columnwidth]{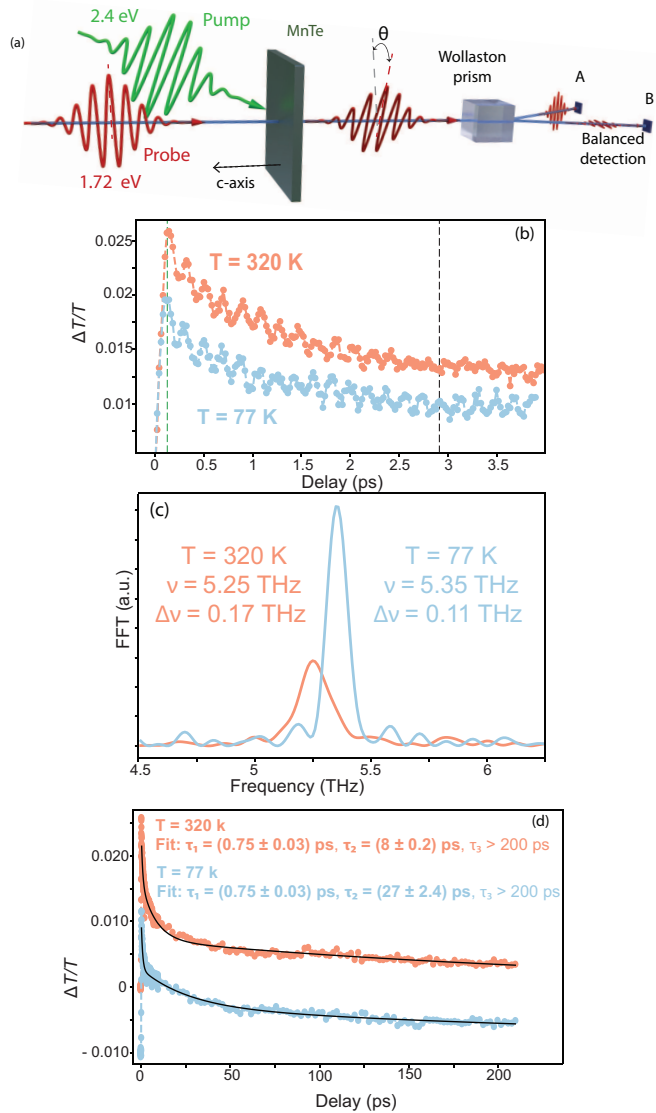}
	\caption{\footnotesize{Temperature-dependent optical dynamics. (a) Set-up to measure the transient transmissivity and rotation of the polarization. The sum of the signals from the two diodes (A+B) is a measurement of $\Delta T/T$, while $\theta$ is given by the difference (A-B) in the balanced detection scheme. The selected probe photon energy corresponds to the maximum signal in Fig. \ref{fig:Fig2}(b), while the pump photon energy is the same as in the experiment reported in Fig. \ref{fig:Fig2}. (b) Transient transmissivity measured in the paramagnetic (T = 320 K) and antiferromagnetic (T = 77 K) phases. (c) Fourier transform of the data in panel (b). (d) Transient transmissivity in the paramagnetic and antiferromagnetic phase detected in a 200 ps delay range.}}
	\label{fig:Fig4}
\end{figure}

We directly detect the photo-induced optical response of \sample\ by tracking the time-resolved transmissivity with a narrow-band, i.e. \textit{single-colour}, optical probe beam. The photon energy of the probe beam was set to 1.72 eV, since at this photon energy a strong signal was observed in Fig. \ref{fig:Fig2}(b). The set-up employed, described in Sec. \ref{sec:Methods} and shown in (Fig. \ref{fig:Fig4}(a)), allows the simultaneous detection of the transient transmissivity ($\Delta T/T$) and the rotation of the polarization. In this Section we discuss the $\Delta T/T$ dynamics.

The data taken in the first 3.5 picoseconds after the photo-excitation (Fig. \ref{fig:Fig4}(b)) show that the 5.3 THz phonon is induced and detected in both the paramagnetic and the antiferromagnetic phases. In particular, the initial phase of the oscillation is cosine-like in both traces, strongly suggesting that the lattice modes are triggered via a displacive mechanism~\cite{Zeiger1992,Merlin1997a}. The essence of this mechanism is that  the excitation of electrons from an occupied to an unoccupied state alters the inter-ionic potentials so that the previous atomic positions are no longer equilibrium positions. The ions thus begin moving towards the minima of the modified potential, at the frequency determined by the dispersion relation of phonons. The excitation photon-energy employed in our experiments is above $E_{\text{gap}}$, so the pump beam actively promotes electrons in the initially unoccupied conduction band triggering the displacive phonon excitations.

Figure \ref{fig:Fig4}(b) shows that the oscillations in the two data sets are not in phase anymore at a delay of $\approx$ 3 ps, hinting towards a temperature dependence of the phonon frequency. Hence we consider the Fourier transform of the data in Fig. \ref{fig:Fig4}(c): both the frequency and the linewidth of the lattice mode are affected by temperature. More precisely, the frequency is lower, and the damping time is shorter in the paramagnetic phase, pointing at an intimate connection between the antiferromagnetic order and the $E_{2g}$ phonons. In solids the frequency and lifetime of lattice vibrations may be temperature-dependent, regardless of any coupling to spins. This is indeed the case also for several phonon modes in \sample\ , as reported by temperature-dependent Raman scattering data~\cite{Zhang:2020jl}. However, the same experiment reveals that the temperature dependence of the 5.3 THz lattice mode differs from the one of the other phonons, as the long-range antiferromagnetic order is established (i.e. below $T_{\mbox{\tiny{N}}}$~\cite{Zhang:2020jl}). More in details, the frequency of the 5.3 THz mode increases as the temperature is lowered in a more pronounced way than in the case of the other phonons. The temperature dependence of the linewidth displays a discontinuity in correspondence of the magnetic phase transition. Both features are unique of this specific mode~\cite{Zhang:2020jl}. The measurements reported here confirm the spin-phonon coupling in the frequency domain from the spectral analysis in Fig. \ref{fig:Fig4}(c). Furthermore, the time-evolution of the phase of the oscillations in the two time-traces in Fig. \ref{fig:Fig4}(b) expresses the consequences of the spin-lattice coupling in the time-domain. In fact, the two curves exhibit oscillations which are out-of-phase at a delay time of 3 ps, as the eigenfrequency of the phonon is temperature-dependent because of the spin-phonon coupling.

Figure \ref{fig:Fig4}(d) shows $\Delta T/T$ dynamics measured over an extended timescale, fitted with a function containing three exponentially decaying terms. The first relaxation has a characteristic time $\tau_1 = (750 \pm 30) $ fs and is temperature-independent. Since the pump beam with 2.4 eV photon energy promotes electrons to the conduction band with excess energy, we associate $\tau_1$ with the relaxation of such carriers via electron-phonon scattering. In contrast, the characteristic time $\tau_2$ is affected by temperature. Interestingly, the variation of $\tau_2$, from 8 ps to 27 ps, matches quantitatively the variation of the recently calculated lattice thermal conductivity~\cite{Mu:2019jg} from 2 W/(m$\cdot$K) at 300 K to 7.5 W/(m$\cdot$K) at 100 K. A first-principle treatment of this problem shows that the phonon thermal conductivity depends on the characteristic phonon-phonon and electron-phonon scattering times~\cite{Tong2019}. Therefore, we associate the time constant $\tau_2$ (see fit in Fig.\ref{fig:Fig4}(d)) with the phonon-phonon scattering time. The data show also a third exponential relaxation $\tau_3$, likely due to heat diffusion from the excited volume. The time constant $\tau_3$ exceeds the observation window and therefore it is not further discussed.

The data discussed so far reports unambiguously the optical activation of coherent phonons. These modes are coupled to the magnetic system and, in the next Section, we derive a model predicting which spin dynamics is triggered by the collective lattice oscillations.

\section{Theory}
\label{sec:theory}

In Sec. \ref{sec:Methods} we have already mentioned that the $E_{2g}$ phonons induce an ionic displacement which modulates the angle between the bonds involved in the Mn-Te-Mn super-exchange pattern (\ref{fig:Fig1}(b)-(c)). A phonon able to induce such an atomic motion has been experimentally demonstrated to couple to the macroscopic magnetic order in Heisenberg antiferromagnets~\cite{tomono1990}. However, a priori it is not trivial to predict how the generation of the 5.3 THz lattice mode affects the spin dynamics. Moreover, the nature of the microscopic coupling between the lattice mode and the spins in \sample \ is an open issue. In the following we derive a model, in which the phonon directly modulates the exchange interaction, and discuss the consequences of such perturbation for the dynamics of the antiferromagnetic order parameter. We emphasize that our model does not rely on quasi-static shifts of the coupling constants due to anharmonicities, as done in other studies~\cite{Fechner:2018,Afanasiev2021}. Microscopically the process we are considering is analogue to a parametric conversion: the phonon drives pairs of magnon modes with equal and opposite wavevector, originating from spin-flip events in different sublattices, thus fulfilling both spin-($\Delta S = 0$) and wavevector-conservation ($\Delta \boldsymbol{k} = 0$). Consequently, pairs of counter-propagating magnons with half the frequency of the phonon are created. From the dispersion relation of spin waves in \sample\ ~\cite{Mu:2019jg}, we conclude that these magnons have a wavevector neither at the centre nor at the edges of the Brillouin zone, but in between. Importantly, no torque on the spins is applied (since $\Delta S = 0$), hence the direction of the sublattice magnetization is unchanged. We model the spin dynamics of \sample\ considering the equilibrium Hamiltonian

\begin{eqnarray}
    H &= &\sum_{\langle i, j \rangle_z} J_1 \vec{S}_i \vec{S}_j + \sum_{\langle i, j \rangle_{xy}} J_2 \vec{S}_i \vec{S}_j
+ \sum_{\langle \langle i, j \rangle \rangle} J_3 \vec{S}_i \vec{S}_j,
  \label{eq:hamiltonian}
\end{eqnarray}

\noindent where $J_i$ denote the Heisenberg couplings between the spins $\vec{S}_i$ and $\vec{S}_j$ with the couplings $J_1$ between interlayer nearest neighbours, $J_2$ between nearest neighbours in a Mn-plane, and $J_3$ between third-nearest neighbours. The effect of the pump pulses on the system is expressed via the excitation of the $E_{2g}$ phonons which, as aforementioned, modulate the exchange interaction. We thus express such perturbation as

\begin{align}
  \frac{\delta J_3(t)}{J_1 S} &\equiv a(t) = a_0 \exp(-\gamma t) \cos(\omega_0 t),
\end{align}

\noindent where $a_0$ parametrizes the maximum value of the change of $J_3$, $\omega_0$ denotes the phonon frequency, and $\gamma$ specifies the damping of the phonon mode. By adding this term to the Hamiltonian in Eq.(\ref{eq:hamiltonian}) and relying on the Heisenberg equation of motion for operators $(\frac{d\hat{A}}{dt} = i \langle [H(t),\hat{A}] \rangle$, where $\hat{A}$ is a generic operator), we can calculate the time dependent order parameter $\vec{L} \equiv$ Mn$^{\Uparrow}$ - Mn$^{\Downarrow}$. Details of the calculations can be found elsewhere~\cite{Deltenre2021}. For the Heisenberg couplings in equillibrium, we use the parameter set of Ref.~\cite{Szuszkiewicz:2006fz}. Note that we have to include a factor of 2 and that we use another sign convention due to a slightly different definition of the Hamiltonian compared to Ref.~\cite{Szuszkiewicz:2006fz}. We represent the spins by bosonic operators via the Holstein-Primakoff representation \cite{PhysRev.58.1098} and keep only bilinear terms in the Hamiltonian (linear spin wave theory). We apply a Fourier and a Bogoliubov transformation to diagonalize the equillibrium problem in $\vec{k}$-space and obtain

\begin{subequations}
  \begin{align}
    \frac{\tilde{H}_0}{J_1 S} &=
    \sum_{\vec{k}} \omega_{\vec{k}} b_{\vec{k}}^\dagger b_{\vec{k}} + E_d + \Delta E \\
    \label{eq:H0-diagonal}
      \text{with } \omega_{\vec{k}} &= \sqrt{A_{\vec{k}}^2 -B_{\vec{k}}^2}\\
      A_{\vec{k}} &=\frac{1}{J_1} \Big(2 J_1 - 6J_2 + 12J_3
      + J_2\gamma_\Delta \left(\vec{k}\right)\Big)\\
    B_{\vec{k}} &= 2 \cos\left(k_z\right) \left(1+ 2 \frac{J_3}{J_1} \gamma_\Delta \left(\vec{k}\right)\right) \\
      \gamma_\Delta\left(\vec{k}\right) &= \cos(k_x) + \cos \left(\frac{\sqrt{3}}{2}k_y + \frac{1}{2} k_x \right) + \\ \nonumber
       &+ \cos\left(-\frac{\sqrt{3}}{2}k_y + \frac{1}{2} k_x \right),
  \end{align}
\end{subequations}

where $\omega_{\vec{k}}$ denotes the magnon frequency; $b_{\vec{k}}^\dagger b_{\vec{k}}$ stands for the magnon occupation operator. The contributions to the ground state energy $E_d$ and $\Delta E$ arise from the Fourier and Bogoliubov transformation, respectively. Since they do not enter into the equations of motion that we calculate in the next step, further treatment is not necessary at this point.
Considering that the Heisenberg couplings $J_i$ oscillate, the Hamiltonian becomes time dependent. Via the Heisenberg equation, we are able to calculate the magnon dynamics
  
\begin{subequations}
	\begin{align}
        \frac{d \langle b_{\vec{k}}^\dagger b_{\vec{k}}\rangle}{dt} &=\frac{d u_{\vec{k}}}{dt} = 2 a(t) \beta_{\vec{k}} w_{\vec{k}} -\gamma_r u_{\vec{k}} \label{eq:eom_u} \\
        \frac{d \text{Re}\, \langle b_{\vec{k}}^\dagger b_{-\vec{k}}^\dagger \rangle}{dt} &=\frac{d v_{\vec{k}}}{dt} = -2(\omega_{\vec{k}} + a(t) \alpha_{\vec{k}}) w_{\vec{k}}\label{eq:eom_v} -\gamma_r v_{\vec{k}} \\
        \frac{d \text{Im}\, \langle b_{\vec{k}}^\dagger b_{-\vec{k}}^\dagger \rangle}{dt} &=\frac{d w_{\vec{k}}}{dt} = 2(\omega_{\vec{k}} + a(t) \alpha_{\vec{k}}) v_{\vec{k}} + \\ \nonumber
        & 2 a(t)  \beta_{\vec{k}} (u_{\vec{k}} +1/2) -\gamma_r w_{\vec{k}}
      \label{eq:eom_w}\ ,
    \end{align}
\label{eq:eom_uvw}
\end{subequations}

where $\alpha_{\vec{k}}$ and $\beta_{\vec{k}}$ depend on which couplings are time dependent. In addition, we include a phenomenological decay rate $\gamma_r$ in analogy to what has been done
for triplone \cite{Yarmohammadi2021}.
We solve the coupled differential equations for discretized $k_z$ and discretized
DOS of the triangular lattice \cite{doi:10.1002/andp.19955070405} numerically using a Bulirsch-Stoer algorithm \cite{10.5555/1403886} with \num{1000} time steps per picosecond. In our description, the sublattice magnetization consists of a time-independent part $L_0 = S -\Delta L$ that contains the spin per site reduced by quantum fluctuations $\Delta L = \frac{1}{2N}
      \sum_{\vec{k}} \left (\omega_{\vec{k}}/ A_{\vec{k}} -1 \right)$ and a time-dependent part $\delta L$: $L = L_0 - \delta L$. The time-dependent change of the sublattice magnetization reads
      
\begin{eqnarray}
    \delta L&=&
    \frac{1}{N}
    \sum_{\vec{k}} \left[\frac{A_{\vec{k}}}{\omega_{\vec{k}}} \langle b_{\vec{k}}^\dagger b_{\vec{k}}\rangle - \frac{B_{\vec{k}}}{\omega_{\vec{k}}} \text{Re}\, \langle b_{\vec{k}}^\dagger b_{-\vec{k}}^\dagger \rangle \right] \label{eq:delta-L-def}.
\end{eqnarray}

Physical meaningful results require that the sublattice magnetization is not greater than $S$. Therefore, we restrict ourselves to small $|\delta L|$, but $\delta L$ may be negative representing an increase of $L$ for short time intervals as along as $\Delta L + \delta L \ge 0$.

\begin{figure}[h]
	\centering
	\includegraphics[width=\columnwidth]{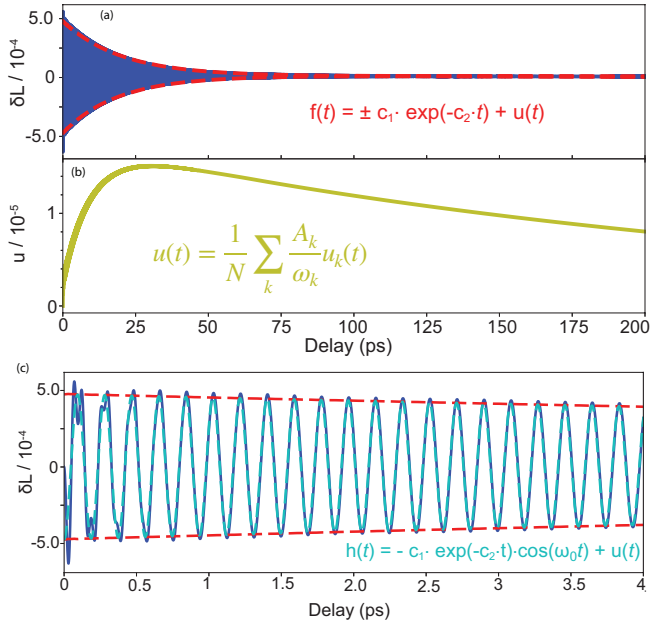}
	\caption{\footnotesize{Spin-lattice coupling. Predicted non-equilibrium dynamics of the $L$ vector on the 200 ps (a) and 4 ps (c) timescale. The increase of the magnon population ($u(t)$) is shown in panel (b). The function $h(t)$ is a fit to $\delta L$. The function $f(t)$ fits the envelope of the spin dynamics, obtained for $c_1 = (4.74\pm 0.04)\cdot 10^{-4}$ and $c_2 = (0.0515\pm0.0007)$ ps$^{-1}$.}}
	\label{fig:Theory}
\end{figure}

Inserting the experimental values of the parameters, the equations are solved and the time-dependence of the change of the longitudinal component of $\vec{L}$ ($\delta L$) is plotted in Fig. \ref{fig:Theory}(a),(c). The purely longitudinal dynamics of the order parameter exhibits oscillations with the same frequency of the driving phonons (Fig.\ref{fig:Theory}(c)). It is worth noting that this constitutes a novel microscopic mechanism to control magnetic systems which has been previously proposed only for spin ensembles lacking a long-range order~\cite{Yarmohammadi2021}. \revisionKB{In fact, the recently proposed nonlinear phononics applied to magnetic materials is based on the idea that resonantly driven infrared-active phonons couple to Raman-active phonons, which in turn modify the exchange integral~\cite{Fechner:2018}. In our case Raman-active phonon modes induce pairs of coherent magnon modes, in a parametric process which fulfills the energy- and wavevector-conservation laws. Uniquely to our mechanism, the dynamics is purely longitudinal as the total spin is conserved, i.e. $\Delta S = 0$. The coherent magnons dephase because many magnon pairs, not only the pair resonant with the frequency matching the phonon, contribute to $\delta L$, although not resonantly and thus with a smaller amplitude. Dephasing is thus fully captured by our theory. Additional possible scattering events are taken into account at a phenomenological level by the $\gamma_r$ parameter. The dephasing of the coherent oscillations also triggers incoherent longitudinal dynamics of the order parameter}. The harmonic component of the signal decays with a rate proportional to the damping of the driving lattice mode (Fig. \ref{fig:Theory}(a)-(c)). Interestingly, the population of a specific magnon mode, expressed by $u_{\vec{k}}$, grows (Fig. \ref{fig:Theory}(b)) effectively demagnetising both sublattices simultaneously. We emphasize that this result is not trivial: typically light-induced spin dynamics is described by means of an additional source term in the Landau-Lifschitz equations of motion, which mimics the action of the optical excitation on spins. Our approach is inherently different. We modify the Hamiltonian of the magnetic system by means of a term describing how the most fundamental magnetic interaction is affected by laser pulses (via the generation of phonons). Then we solve the quantum mechanical Heisenberg equation of motion, obtaining the dynamics of the antiferromagnetic order parameter.

As already discussed in Sec. \ref{sec:Methods}, our experimental geometry is sensitive to magneto-optical effects, which are in principle able to reveal the spin dynamics predicted by means of our model. The magneto-optical measurements are discussed in the next section. 

\section{Single-color magneto-optical pump-probe measurements}
\label{sec:1ColMO}

\begin{figure}[t!]
	\centering
	\includegraphics[width=\columnwidth]{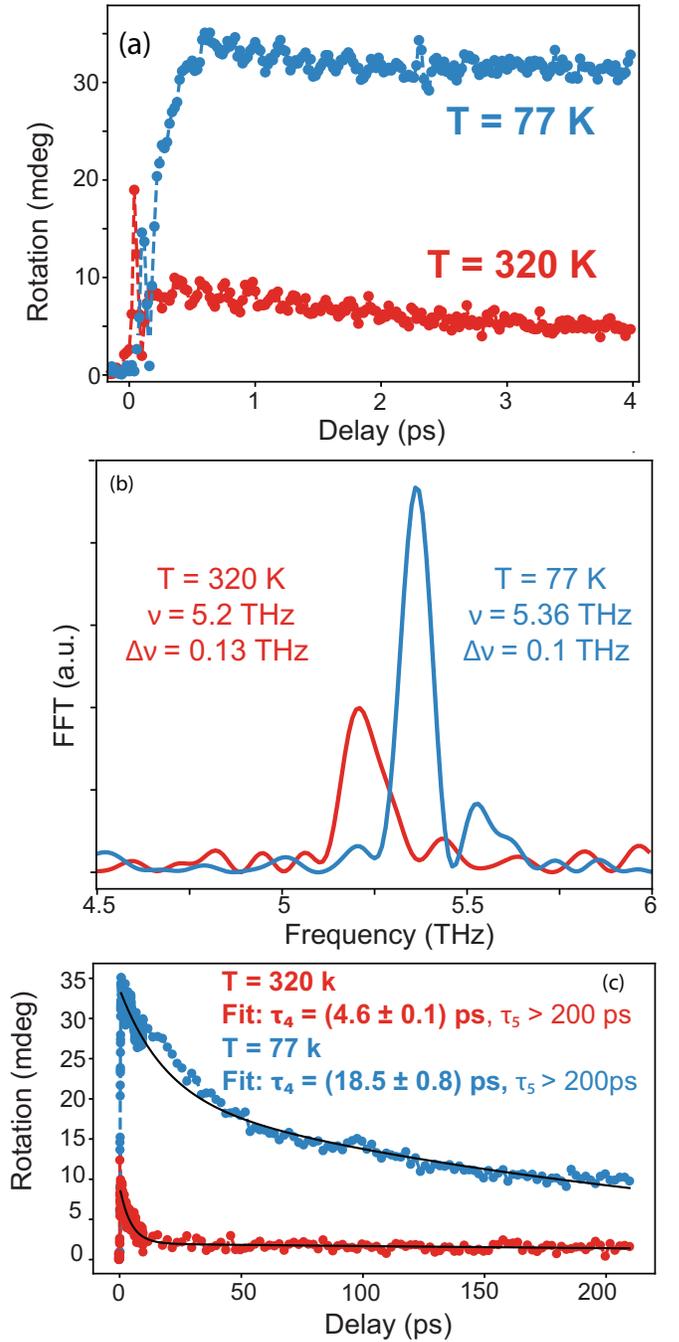}
	\caption{\footnotesize{(a) Rotation of the polarization in the paramagnetic and antiferromagnetic phase in a 4 ps temporal range.(b) Fourier transform of the data in panel (a). (c) Rotation of the polarization in the paramagnetic and antiferromagnetic phase is shown in a 200 ps temporal region.}}
	\label{fig:Fig5}
\end{figure}
	
The magneto-optical measurements are performed with the set-up described in Sec. \ref{sec:Methods} and shown in (Fig. \ref{fig:Fig4}(a)). The time-dependent measurements of the rotation of the polarization $\theta$ shown in Fig. \ref{fig:Fig5}(a)-(c) are performed in the same experimental conditions employed for collecting the data in Fig.\ref{fig:Fig4}(b),(d), so that a direct comparison with the $\Delta T/T$ dynamics can be carried on.  In the first four picoseconds subsequent the photo-excitation, $\theta$ oscillates at the 5.3 THz phonon frequency both in the paramagnetic and in the antiferromagnetic phase. The Fourier transform of the data reported in Fig. \ref{fig:Fig5}(b) confirms the observations discussed in Fig. \ref{fig:Fig4}(c) and coupling of phonons to the magnetic system. The dynamics of $\theta$ in a 200 picosecond time window (Fig. \ref{fig:Fig5}(c)) departs from $\Delta T/T$. The femtosecond relaxation dynamics $\tau_1$, which is ascribed to electron-phonon interaction, is not present. A temperature-dependent picosecond $\tau_4$ and a much longer dynamics $\tau_5$, which is not completely exhausted in the 200 ps range here analysed, are observed.

We compare the $\Delta T/T$ and $\theta$ measurements, by expressing the interaction between \sample\ and the probe beam in terms of the dielectric tensor (see Sec. \ref{sec:Methods}). The rotation of the polarisation cannot express a magneto-optical effect linear in the antiferromagnetic vector, since an antiferromagnetic counterpart of the Faraday effect is symmetry-forbidden in a centro-symmetric material~\cite{Krichevtsov1993} such as \sample\ . The symmetry analysis of \sample\ reported in Sec. \ref{sec:Methods} points at active quadratic magneto-optical effects, common in antiferromagnets~\cite{Bossini:2017bo,Nemec:2018gw}, such as magnetic linear dichroism. This effect is expressed by the rotation of the polarization of the probe, which depends on the order parameter via contributions proportional to $L \Delta L$. This is consistent with the measurements of ultrafast longitudinal and transversal dynamics of the N\'eel vector via second order magneto-optics~\cite{Bossini2014,Bossini:2017bo,10.1038/nphoton.2016.255,Nemec:2018gw}.

\revisionKBR{Following this discussion it is thus tempting to ascribe the origin of the oscillations detected in Fig. \ref{fig:Fig5}(a) to coherent spin dynamics. Although this effect is predicted by the model and its detection is symmetry-allowed in our experimental configuration, we cannot neglect that a non-magnetic contribution to the rotation in Fig. \ref{fig:Fig5}(a) ($\Im{\epsilon_{ii}^0}$ in Eq.(\ref{eq:ep}) is present for sure, since the 5.3 THz phononic oscillations are observed also above the ordering temperature. Moreover, the predicted coherent magnetic oscillations have the same frequency as the phonon mode, as a result of the parametric process. A quantitative decoupling of the lattice and spin contributions to the oscillations requires additional measurements. Differently, the incoherent dynamics can be straightforwardly ascribed to spin dynamics.}

The incoherent contributions to the signal is considered by comparing the measurements performed in the paramagnetic and antiferromagnetic phases of \sample\ . The amplitude of the rotation signal (Fig. \ref{fig:Fig5}(a)) is significantly affected, decreasing by a factor of three in the paramagnetic phase, while this is not the case for $\Delta T/T$ (Fig. \ref{fig:Fig4}(b)). This is an evidence of the magnetic contributions to the rotation measurements. Moreover, the rotation detected in a 200 ps time window (Fig. \ref{fig:Fig5}(c)) deviates from $\Delta T/T$ as well. In particular, $\tau_4$ increases from $\tau_4 = (4.6 \pm 0.1)$ ps to $\tau_4 = (18.5 \pm 0.8)$ ps when the long-range order is established. These values with their errorbars are different from the time constants observed in the optical measurements (Fig. \ref{fig:Fig4}(d)). From the linewidth of the lattice mode (Fig. \ref{fig:Fig4}(c)), we estimate its lifetime to be $\approx 10$ ps, which is on the same order of magnitude of the fitted value of $\tau_4$ in the antiferromagnetic phase. Our model predicts that the lifetime of the 5.3 THz phonon determines the time-scale of the sublattice demagnetization, i.e. the envelope of the harmonic signal shown in Fig.\ref{fig:Theory}(a),(c), which can be measured via magnetic linear dichroism~\cite{Bossini2014,Bossini:2017bo,10.1038/nphoton.2016.255,Nemec:2018gw}. \revisionKBR{ We thus ascribe $\tau_4$ to the sublattice demagnetization. The blue curve in Fig. \ref{fig:Fig5}(a) exhibits a contribution at a sub-picosecond timescale. We do not interpret this component of the signal as ultrafast demagnetization as, unlike metals, a semiconductor does not possess the free electrons necessary to photo-induce the femtosecond quench of the magnetic order. To the best of our knowledge, a femtosecond demagnetisation of a dielectric or semiconductor is yet to be reported. As a matter of fact, experimental observations in magnetic semiconductors demonstrate that the first picosecond of laser-driven spin dynamics can display even an enhancement of the magnetization~\cite{Liu2012}, instead of the demagnetization ubiquitous in metals~\cite{Koopmans2010,Kirilyuk2010}. Identifying the origin of the sub-picosecond contribution to the signal goes beyond the scope of the present paper. From the results of our model, we expect the $\tau_4$ demagnetization to be driven by the photo-induced phonons. Nevertheless, we note that also electronic effects could play a role, as the pump photon-energy is tuned above the band-gap. This contribution should not be dominant in our experiment, since the timescale of the demagnetization ($\tau_4$) does not match the timescale of the electronic relaxation observed in the transient transmissivity ($\tau_1$). However we cannot quantitatively disentangle the electronic and lattice contributions to the signal.}

\revisionKB{We observe that our sample is in a multi-domain state. It may thus be expected that the magneto-optical signal should be averaged out. However this is not the case: recent theoretical and experimental results~\cite{Bossini2021NiO,Gomonay2021} show that a multi-domain antiferromagnet can exhibit both coherent and incoherent spin dynamics. In fact, the efficiency of the optical excitation (for both coherent and incoherent processes) differs for different domains, as it is defined by the direction of spins with respect to the polarization of light. The same argument applies to the magneto-optical effects employed to monitor spin dynamics. Consequently, the contribution of some domains is dominant and hence a net signal can be detected.}

The idea that optically-excited phonons can induce spin dynamics via a modification of the exchange interaction was put forward in the case of dielectric ferrimagnet~\cite{Maehrlein:2018iq} and weak-ferromagnet~\cite{Afanasiev2021}. However, our approach and our work are essentially different. In the former case~\cite{Maehrlein:2018iq}, the optically driven phonons decay via scattering with other lattice modes, so that the lattice itself is a laser-heated thermal bath, with temperature higher than the magnetic system. In this context, stochastic fluctuations of the ionic positions affect the exchange interaction, which in turn triggers the demagnetization of the sublattices. \revisionKB{In the latter case~\cite{Afanasiev2021}, the resonant pumping of phonons results in the generation of coherent magnons at the center of the Brillouin zone, with a modified frequency in comparison with the dispersion relation}. 
Our model describes the perturbation of the exchange by means of a specific coherent lattice mode. The corresponding microscopic mechanism is totally different, as it entails a phonon-driven parametric generation of counter-propagating magnon pairs, not at the center of the Brillouin zone. Additionally also the spin dynamics calculated in the present work is unique since the coherent harmonic component of $L$ shown in Fig. \ref{fig:Theory}(c) does not appear in the literature~\cite{Maehrlein:2018iq}. \revisionKB{The oscillating signal detected in Ref.~\cite{Afanasiev2021} correspond to mascroscopic spin precession of the magnetization of a weak-ferromagnet and not to the solely longitudinal coherent motion of $L$ described in our model. We finally stress again, that Ref.~\cite{Afanasiev2021} relies on static, average shifts of the exchange couplings, while we do not need to take this effect into account.}

\section{Conclusion}
\label{sec:Concl}

The essence of spintronics consists in the coupling between charges and spins. Here, we demonstrate that in a room temperature antiferromagnetic semiconductor optical phonons couple dynamically to both spin and charges at the ultrafast time-scale. It can thus be envisaged that a resonant pumping of the proper phonon mode, which is symmetry-allowed in the case of \sample\ ~\cite{Cyvin1965}, represents a channel to control both spin and charges in semiconductors even in a nonlinear dynamical fashion, a cornerstone for ultrafast semiconductor spintronics. The concept described in our paper could be extended to atomically-thin materials, as an efficient displacive phonon-pumping has already been reported for these systems~\cite{Trovi2020}.

\section*{Acknowledgments}
{\footnotesize{This work was supported by the Deutsche Forschungsgemeinschaft through the International Collaborative Research Centre 160 (Projects No. B8 and No. Z4) and by UH 90/14-1,
by the COST Action MAGNETOFON (Grants No. CA17123, STSM No. CA17123-44749). S.C.D. acknowledges support by the PRIN2017 project Central (ID 20172H2SC4). D.B. acknowledges support from the Deutsche Forschungsgemeinschaft (DFG) program BO 5074/1-1. G.S. Acknowledges Support by the Austrian Science Fund (FWF), Projects No. P30960-N27 and I 4493-N. A.B. acknowledges Austrian Science Fund (FWF), Projects No. P26830 and No. P31423.}}


%

\end{document}